\documentclass{elsart} 
\usepackage{amsmath,amssymb}
\usepackage{graphicx}
\usepackage{multirow}

\begin{document}

\begin{frontmatter}

\title{Complex Langevin Equations and Schwinger-Dyson Equations}

\author{Gerald Guralnik}
\ead{gerry@het.brown.edu},
\author{Cengiz Pehlevan\corauthref{cor}}
\corauth[cor]{Corresponding author.}
\ead{cengiz@het.brown.edu}

\address{Department of Physics, Brown University, Providence, RI 02906 USA}

\begin{abstract}  
  Stationary distributions of complex Langevin equations are shown to be the
  complexified path integral solutions of the Schwinger-Dyson equations of the
  associated quantum field theory.  Specific examples in zero dimensions and
  on a lattice are given. Relevance to the study of quantum field theory phase
  space is discussed.
\end{abstract}


\end{frontmatter}

\section{Introduction}

One  frequently comes across complex valued path integral weights in physics. A
major example is the study of field theories in Minkowski space. Even in
Euclidean field theories the effective action may turn out to be complex,
e.g. QCD with a chemical potential. Another case, in which we will be
interested in this paper, is the study of the solution space of
Schwinger-Dyson equations as the phase space of the associated quantum field
theory \cite{Garcia96,Guralnik07}.

Standard Monte Carlo methods do not work in the case, where the path integral
weight is not positive-definite. This is the famous ``sign problem''. The root
of the problem lies in the fact that Monte Carlo methods work on the
probabilistic interpretation of the path integral weight $e^{-S}$, where $S$
is the action of the system. In the examples above, the exponent becomes
complex and the probabilistic interpretation fails.

Complex Langevin equations were proposed by Parisi \cite{Parisi83} and Klauder
\cite{Klauder83} to simulate systems with complex valued path integral
weights. For $d$ dimensional systems with real weights, a Langevin equation in
$d+1$ dimensions may be used to study the partition function of the
system. When properly set up, the Langevin process converges to a unique
stationary distribution, which is the partition function of the associated
system, in the limit of the additional dimension going to infinity. This fact
was used by Parisi and Wu in the stochastic quantization of quantum fields
\cite{Parisi81}. For systems with complex actions, one can still write down a
(complex) Langevin equation, as suggested by Parisi and Klauder, but this
approach comes with many problems. First of all, it is not certain that the
complex Langevin simulation will ever converge to a stationary distribution
and if it does, there may be many such stationary distributions, see
e.g. \cite{Gausterer93,Gausterer94,Gausterer98,Salcedo93,Lee94}. Salcedo noted
that these stationary distributions may be constructed by path integrals over
contours that connect zeros of the path integral weight $e^{-S}$
\cite{Salcedo93}. Other authors noted that these stationary distributions
satisfy Schwinger-Dyson equations, e.g. \cite
{Berges07,Aldazabal85,Ito93,Xue86}. On a completely different track of
research, one of the authors with Garcia and Guralnik studied the solution
space of Schwinger-Dyson equations and noted that different solutions to
Schwinger-Dyson equations may be written as path integrals over contours that
connect zeros of the path integral weight $e^{-S}$ \cite{Garcia96,
  Guralnik07}, exactly as Salcedo suggested for stationary distributions of
the complex Langevin equation. Furthermore, the authors of references
\cite{Garcia96} and \cite{Guralnik07} suggested different solutions of the
 Schwinger-Dyson equations be interpreted as different phases of the
 associated quantum field theory and studied this proposal in
detail. Salcedo in his paper \cite{Salcedo93} made the same suggestion about
stationary distributions of the complex Langevin equations in one sentence at
the conclusion, but did not elaborate on it. Our aim in this paper is to point
out and clarify the connection between these two lines of research and propose
@@ complex Langevin equations as a basis of a new numerical method of studying different phases
of a quantum field theory.  \indent We start by a rather detailed explanation
of the mechanism of complex Langevin equations and the problems associated
with them in section \ref{Langevin}. Section \ref{stationary} contains the
main result of this paper. There, in a zero dimensional setting, we show that
the stationary distributions of a complex Langevin equation are the solutions
of the Schwinger-Dyson equations for the associated quantum field
theory. Furthermore, these solutions may be constructed by changing the
integration contour of path integrals from real paths to contours that connect
the zeros of the path integral weight $e^{-S}$ on the complex plane. In
section \ref{lattice}, we do the trivial generalization of the problem to a
lattice and discuss related issues. We conclude by further summarizing the
results of \cite{Garcia96} and \cite{Guralnik07} and point out the connection
between different phases of a quantum field theory and stationary
distributions of complex Langevin equations. Based on this observation, we
propose complex Langevin equations as a numerical method of studying different
phases of a quantum field theory.

\section{The Complex Langevin Equation}\label{Langevin}

In this section, we introduce Langevin equations. We discuss zero dimensional
quantum field theories with a scalar field for simplicity. Generalization to
vector fields and higher dimensions is straightforward, see
e.g. \cite{Namiki92}. For systems with action $S\left(\phi\right)$, we are
interested in calculating expectation values like
\begin{equation}\label{Exp}
  \left<\begin{cal}O\end{cal}\right>=\frac{\int d\phi \, \begin{cal}O\end{cal}(\phi) e^{-S(\phi)} }{\int d\phi \,e^{-S(\phi)}},
\end{equation} 
using Langevin equations. We will assume $S\left(z\right)$ to be an analytic function of the complex variable $z$.

Note that our formulation is a Euclidean space formulation, i.e. the path
integral weight will be complex only when the action is complex. However in
Minkowski space formulations of quantum field theories, the path integral
weight will still be complex even though the action is real, since the weight
is defined by $e^{iS_M(\phi)}$.  When we speak about complex actions in
Euclidean space, our results will be applicable to Minkowski space actions
after necessary modifications. One has to note that $S(\phi) = -iS_M(\phi)$
and introduce appropriate factors of $i$'s in the generating function
definition and terms of Schwinger-Dyson equation.

\subsection{Real Actions}

When the action is real (for the moment $\phi$ is a real field), one can
create a stochastic process using a Langevin equation with a unique stationary
distribution $e^{-S(\phi)} / \int d\phi\, e^{-S(\phi)}$:
\begin{equation}\label{LangevinEq}
  d\phi(\tau) = -\frac{\partial S}{\partial\phi(\tau)}d\tau + dw(\tau),
\end{equation}
where $\tau$ is a fictitious time and $w\left(\tau\right)$ is the Wiener process normalized to satisfy:
\begin{align}\label{noise}
  \left<dw(\tau)\right>=0, \qquad \left<dw(\tau)dw(\tau)\right>=2d\tau, \qquad \left<dw(\tau_1)dw(\tau_2)\right> = 0 \quad (\tau_1 \neq \tau_2).
\end{align}
Then one can run this Langevin process to calculate the intended expectation
values as in equation (\ref{Exp}). We first show that this Langevin process
really converges to the intended stationary distribution.

Associated with the Langevin process is a probability density $P(\phi,\tau)$
\begin{equation}
  \left<F(\phi\left(\tau\right))\right> = \int d\phi \,F(\phi)P(\phi,\tau),
\end{equation}
which can be shown (e.g. \cite{Gardiner04}) to satisfy the Fokker-Plack equation
\begin{equation}\label{Fokker_Planck}
  \frac{\partial P(\phi,\tau)}{\partial\tau} = \frac{\partial}{\partial \phi}\left(\frac{\partial}{\partial \phi}+\frac{\partial S}{\partial\phi}\right) P(\phi,\tau). 
\end{equation}
Note that the Fokker-Planck equation enables the normalization condition
\begin{equation}\label{normalization}
  \int d\phi\, P(\phi,\tau) = 1
\end{equation}
to be independent of time since
\begin{align}
  \frac d{d\tau}\int d\phi\, P&(\phi,\tau) = \int d\phi\, \frac{\partial}{\partial \phi}\left[\left(\frac{\partial}{\partial \phi}+\frac{\partial S}{\partial\phi}\right) P\left(\phi,\tau\right)\right] = 0,
\end{align}
with an appropriate boundary condition on $P(\phi,\tau)$.

Now we look at the asymptotic behavior of $P(\phi,\tau)$ as fictitous time goes to infinity. Introducing the quantity
\begin{equation}
  Q\left(\phi,\tau\right) \equiv P(\phi,\tau)e^{S(\phi)/2},
\end{equation}
we can rewrite equation (\ref{Fokker_Planck}) as
\begin{align}\label{FP_Q}
  \frac{\partial Q(\phi,\tau)}{d\tau} &= -\left(-\frac{\partial}{\partial\phi}+\frac{1}{2}\frac{\partial S}{\partial \phi}\right)\left(\frac{\partial}{\partial\phi}+\frac{1}{2}\frac{\partial S}{\partial \phi}\right)Q(\phi,\tau) \nonumber \\
  &= -\left[-\frac{\partial^2}{\partial \phi^2} -\frac{1}{2}\frac{\partial^2S}{\partial\phi^2}+\frac{1}{4}\left(\frac{\partial S}{\partial\phi}\right)^2 \right]Q(\phi,\tau)\nonumber \\
  &= -H_{FP} \, Q(\phi,\tau) . 
\end{align}
where the Hermitian operator $H_{FP}$ is called the Fokker-Planck Hamiltonian. Then
\begin{equation}
  Q_0(\phi)= e^{-S(\phi)/2}
\end{equation}
is an eigenstate of the Hamiltonian $H_{FP}$ with eigenvalue $0$. Furthermore, (assuming $Q_0\in L^2 $) it is the ground state since it is nowhere vanishing. Then the time independent eigenvalue equation
\begin{equation}
  H_{FP}Q_n(\phi) = E_n Q_n(\phi),
\end{equation} 
has solutions with the property $E_n>0$, for $n>0$. Using these eigenvalues and eigenfunctions, one can write any solution to equation (\ref{FP_Q}) as
\begin{equation}
  Q(\phi,\tau)= P(\phi,\tau)e^{S(\phi)/2}=\sum_{n=0}^{\infty}a_n\,Q_n(\phi)e^{-E_n\tau}.
\end{equation}
Since we are looking for solutions $P(\phi,\tau)$ that are probability distributions, we apply the normalization condition (\ref{normalization}), which sets $a_0 = 1/\int d\phi \, e^{-S(\phi)}$. This result is obtained by using the orthonormality of the eigenfunctions $Q_n$. Other coefficients are set by the initial probability distribution associated with the random variable $\phi(\tau)$ at $\tau = 0$. Then the limit 
\begin{equation}
  \lim_{\tau\rightarrow\infty} Q(\phi,\tau) = a_0\, Q_0(\phi),
\end{equation}
implies that one recovers the desired stationary probability distribution for the Langevin process as the fictitous time goes to infinity:
\begin{equation}
  \lim_{\tau\rightarrow\infty} P(\phi,\tau) = \frac{e^{-S(\phi)}}{\int d\phi \, e^{-S(\phi)}}.
\end{equation}
Note that this result is independent of the initial conditions. Going back to the expectation value problem (\ref{Exp}),
\begin{equation}
  \lim_{\tau\rightarrow\infty} \left<\begin{cal}O\end{cal}(\phi(\tau))\right> = \left<\begin{cal}O(\phi)\end{cal}\right>,
\end{equation}
with $\left<\begin{cal}O(\phi)\end{cal}\right>$ given by \eqref{Exp} and ergodicity assures the averaging over the path prescription 
\begin{equation}
  \left<\begin{cal}O(\phi)\end{cal}\right> = \lim_{T\rightarrow\infty}\frac{1}{T}\int_0^T\begin{cal}O\end{cal}(\phi(\tau)) d\tau.
\end{equation}

These observations have been used by Parisi and Wu in the past to formulate the stochastic quantization of quantum fields \cite{Parisi81}. See e.g. \cite{Damgaard87,Namiki92,Namiki93} for a detailed discussion.

\subsection{Complex Actions}

Now we turn to the case where the action is complex. We want to know if we can still use the Langevin equation to calculate desired expectation values. We start by rewriting equation (\ref{LangevinEq}) (now $S$ being complex) in terms of two real variables $\phi_R(\tau)$ and $\phi_I(\tau)$ as
\begin{align}\label{ComplexLangevin}
  d\phi_R(\tau) &= -Re\left[\frac{\partial S}{\partial \phi(\tau)}\right]d\tau + dw(\tau), \nonumber \\  
  d\phi_I(\tau) &= -Im\left[\frac{\partial S}{\partial \phi(\tau)}\right],
\end{align}
where $\phi(\tau) =\phi_R(\tau) + i\phi_I(\tau)$. $w(\tau)$ is again the Wiener process that is normalized to satisfy the mean and variance conditions of equation (\ref{noise}). Note that the equation for $d\phi_I$ has a zero diffusion coefficient (see \cite{Nakazato86} for an example where it is not zero), but still is a stochastic equation through its dependence on $\phi_R$. Complex Langevin equations may be modified to include a term, the kernel, that may be useful to stabilize the system, e.g \cite{Soderberg88, Okamoto89}. For our purposes, we focus on equation (\ref{ComplexLangevin}). Note that we have two different random variables $\phi_R(\tau)$ and $\phi_I(\tau)$, therefore the real probability distribution associated with equation (\ref{ComplexLangevin}) will be of the form
\begin{equation}\label{P}
  \left<F(\phi(\tau))\right> = \int d\phi_R d\phi_I \,F(\phi_R + i\phi_I) P(\phi_R,\phi_I,\tau)
\end{equation}
where we assume $F$ to be analytic. 

There are two important questions related to this process. The first question is whether the probability distribution $P(\phi_R,\phi_I,\tau)$ converges to a stationary distribution at all,
\begin{equation}
  \lim_{\tau\rightarrow\infty}P(\phi_R,\phi_I,\tau) \stackrel{?}{=} \hat{P}(\phi_R,\phi_I). 
\end{equation}
If it does, does it converge to the desired result:
\begin{align}
  \int d\phi_R d\phi_I \,F(\phi_R + i\phi_I)&\hat{P}(\phi_R,\phi_I) \stackrel{?}{=} \frac{\int d\phi_R \,F(\phi_R) e^{-S(\phi_R)}}{\int d\phi_R \, e^{-S(\phi_R)}}.
\end{align}
None of these questions have been completely answered so far. Some rigorous conditions to verify aposteriori the correct convergence of the process are given in \cite{Gausterer93,Gausterer94,Gausterer98}. 

To understand the difficulties related to the convergence problem, we derive the Fokker-Planck equation. First we note that applying the rules of Ito calculus (see for example \cite{Gardiner04}) to the complex Langevin equation (\ref{ComplexLangevin}) will give the identity
\begin{align}\label{main}
\frac d{d\tau}\left<F(\phi(\tau)) \,\right> &= \left<\frac{\partial^2F}{\partial \phi_R^2} - \frac{\partial F}{\partial \phi_R}Re\left[\frac{\partial S}{\partial \phi}\right] - \frac{\partial F}{\partial \phi_I}Im\left[\frac{\partial S}{\partial \phi}\right] \right> \nonumber \\
& = \left<\frac{\partial^2 F}{\partial \phi^2} - \frac{\partial F}{\partial \phi}\left[\frac{\partial S}{\partial \phi}\right]\right>,
\end{align}
where the last line follows from the analyticity of $F(\phi)$. Then using equation (\ref{P}) one can show that $P(\phi_R,\phi_I,\tau)$ (assuming appropriate differentiability and boundary conditions) satisfies the following Fokker-Planck equation:
\begin{align}\label{F_FP}
  \frac{\partial P(\phi_R,\phi_I,\tau)}{\partial \tau} &=  O_{FP}P(\phi_R,\phi_I,\tau) \nonumber \\  &=\left(\frac{\partial^2}{\partial \phi_R^2} + \frac{\partial}{\partial \phi_R}Re\left[\frac{\partial S}{\partial \phi}\right]+ \frac{\partial}{\partial \phi_I}Im\left[\frac{\partial S}{\partial \phi}\right] \right) P(\phi_R,\phi_I,\tau). 
\end{align}
A general statement on the existence of a unique zero eigenvalue (stationary) solution $\hat{P}(\phi_R,\phi_I)$ for the operator $O_{FP}$ cannot be made. Furthermore, zero eigenvalue solutions may exist in the sense of distributions \cite{Gausterer98}. 

One can assume a complex valued function $\tilde{P}(\phi_R,\tau)$ on the real axis such that
\begin{align}\label{tilde}
   \int_{\Re} d\phi_R \,&F(\phi_R) \tilde{P}(\phi_R,\tau) = \int d\phi_R d\phi_I \,F(\phi_R + i\phi_I)\, P(\phi_R,\phi_I,\tau),
\end{align} 
based on the implicit assumption that this equation actually has a solution \cite{Parisi83}. The reverse question, existence of a positive $P(\phi_R,\phi_I,\tau)$ given a complex $\tilde{P}(\phi_R,\tau)$ is disscussed in \cite{Salcedo97,Weingarten02,Salcedo07}. Using this definition, analyticity of $F(\phi)$ and integration by parts in equation (\ref{main}) one can show that $\tilde{P}(\phi_R,\tau)$ satisfies the pseudo Fokker-Planck equation
\begin{align}\label{Fokker_Planck_tilde}
\frac{\partial \tilde{P}(\phi_R,\tau)}{\partial\tau} &= \tilde{O}_{FP}\tilde{P}(\phi_R,\tau) = \frac{\partial}{\partial \phi_R}\left(\frac{\partial}{\partial \phi_R}+\frac{\partial S}{\partial\phi_R}\right) \tilde{P}(\phi_R,\tau),
\end{align}
which has the same form as that of equation (\ref{Fokker_Planck}). 

A formal solution to equation (\ref{tilde}) was introduced in \cite{Nakazato87}. First note that
\begin{equation}
  F(\phi_R+i\phi_I) = e^{i\phi_I\frac{\partial}{\partial \phi_R}}F(\phi_R), 
\end{equation}
due to the analyticity of $F(\phi)$. Inserting this statement into equation (\ref{tilde}) and performing the partial integration assuming necessary boundary and differentiability conditions,
\begin{equation}\label{P_tilde}
 \tilde{P}(\phi_R,\tau) =  \int d\phi_I \, e^{-i\phi_I\frac{\partial}{\partial \phi_R}} P(\phi_R,\phi_I,\tau).
\end{equation}

Because the action $S(\phi_R)$ is complex, one cannot make general statements about the spectrum of $\tilde{O}_{FP}$ (see \cite{Klauder85} for the spectral theorem for a limited class of such operators). Furthermore, the relation between the pseudo Fokker-Planck equation and the complex Langevin equation were derived based on certain assumptions. This should be understood in the sense of distributions, meaning only a formal expression of the identity \eqref{main}. Note that $e^{-S(\phi_R)}$ is still a stationary solution (i.e. $\tilde{O}_{FP}\,e^{-S(\phi_R)} = 0$), but in general the stationary solutions exist as distributions and the uniqueness of stationary solution is not certain \cite{Salcedo93}.

\section{Stationary Distributions of the Complex Langevin Equation and the Boundary Conditions of the Schwinger-Dyson Equation}\label{stationary} 

Despite the difficulties in proving rigorous results, complex Langevin simulations have been used to study many different problems. Interesting cases are those for which the simulation converges to a stationary distribution which is not equivalent to the original complex distribution, e.g. \cite{Parisi83, Lin86, Salcedo93, Hamber85, Lee94, Flower86}. This must be related to the existence of other stationary distributions. Here we discuss a conjecture related to the stationary distributions by Salcedo \cite{Salcedo93} and the well know result that stationary distributions satisfy Schwinger-Dyson identities, e.g. \cite {Berges07, Aldazabal85, Ito93, Xue86}. We show that they follow from one another, based on other work on the boundary conditions of Schwinger-Dyson equations \cite{Garcia96,Guralnik07}. We will point out the results of these references during our discussion.

We start by assuming that the complex Langevin process has a stationary state. For the stationary distribution, the LHS of equation (\ref{main}) must be zero. Then setting $F(\phi) = \phi^n$, $n=1,2,3,\ldots,$ reproduces the Schwinger-Dyson identities for the Green's functions of the quantum field theory defined by the action $S$, i.e.
\begin{align}\label{SDidentity}
&n = 1 \qquad \longrightarrow &&\left<\frac{\partial S}{\partial \phi}\right> = 0,  \nonumber \\
&n = 2 \qquad \longrightarrow &&\left<\phi\frac{\partial S}{\partial \phi}\right> = 1, \nonumber \\
&\ldots, \nonumber \\
&n = k \qquad \longrightarrow &&\left<\phi^{k-1}\frac{\partial S}{\partial \phi}\right> = (k-1)\left<\phi^{k-2}\right>.
\end{align}
Now define the generating function $Z(j)$ for the stationary distribution
\begin{align}\label{Zdef}
  Z(j) = \sum_{n=0}^\infty \frac{\left<\phi^n\right>j^n}{n!} = \left< e^{j\phi} \right>. 
\end{align}
We will assume that the radius of convergence for this series is nonzero. Then there exists a neighbourhood of $j$ around $j=0$ such that the Schwinger-Dyson differential equation holds. i.e.
\begin{equation}\label{SD}
  \left.\frac{\partial S}{\partial \phi}\right|_{\phi = \frac d{dj}}Z(j) = jZ(j).
\end{equation}
To see that \eqref{SD} produces the same identites as \eqref{SDidentity}, substitute the definition of the generating function $Z(j)$, equation \eqref{Zdef}, in the Schwinger-Dyson equation, differentiate with respect to $j$ an appropriate number of times and set $j=0$ at the end. One gets the identities \eqref{SDidentity} order by order at the end of this procedure.

Solutions of the Schwinger-Dyson equation are the stationary distributions of the complex Langevin equation. We solve equation (\ref{SD}) following  \cite{Garcia96}. First define
\begin{equation}
  Z_\Gamma(j) = \int_\Gamma d\phi \, G(\phi) e^{j\phi},
\end{equation}
where $\Gamma$ is a contour over the complex plane. Inserting this into equation (\ref{SD}) one gets:
\begin{equation}\label{contour}
0 = - \left. G(\phi) e^{j\phi}\right|_{\partial\Gamma} + \int_\Gamma d\phi \left[\frac{\partial S}{\partial \phi}G(\phi) + \frac{dG(\phi)}{d\phi}\right]e^{j\phi}. 
\end{equation}
This equation can be solved for 
\begin{equation}
G(\phi) = e^{-S(\phi)},
\end{equation}
and $\Gamma$ is contour that connects the zeros of $e^{-S(\phi)+j\phi}$ on the complex $\phi$ plane. 

\begin{figure}
\begin{center}
\includegraphics[width=80mm]{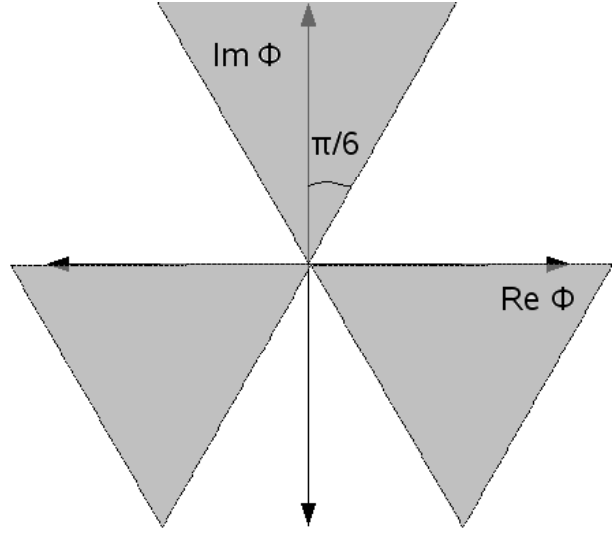}
\end{center}
\caption{The region defined by $\sin(3\theta)<0$, where $\theta$ is the argument of $\phi$, is shaded. Any path starting and ending at infinity within these wedges corresponds to a particular solution of zero dimensional $i\phi^3$ theory. 
\label{FigSectors}}
\end{figure}

Now consider polynomial actions, 
\begin{equation}
S(\phi) = \sum_{i=1}^m \frac 1l g_l\phi^l, \quad m>1.
\end{equation}
The contours will be defined by $m$ wedges, where $m$ is the order of $S(\phi)$, such that $Re(g_m\phi^m)\rightarrow+\infty$ as $\left|\phi\right|\rightarrow\infty$. Contours obtained by deforming $\Gamma$ without crossing singularities of $e^{-S(\phi)}$ and keeping boundary points fixed result in the same generating function. Note that set of all $Z_\Gamma(j)$ will not be independent. The Schwinger-Dyson equation will be of order $(m-1)$, and will have $(m-1)$ independent solutions. For example, if $S=i\phi^3$, then $e^{-S(\phi)}$ will have three zeros in the complex plane, Figure \ref{FigSectors},which will define three different generating functions. However the Schwinger-Dyson equation will be a second order linear differential equation, which has two independent solutions. Therefore any two of the three possible paths will define an independent solution set. 

Since the Schwinger-Dyson equation is linear, any linear combination of the (independent) solutions will also be a solution, 
\begin{equation}
  Z(j) = \sum_{\Gamma_I}a_{\Gamma_I} Z_{\Gamma_I}(j),
\end{equation}
where $\Gamma_I$ define an independent subset of paths $\Gamma$. Now defining the distribution $\tilde{P}_\Gamma(\phi_R)$ on the real plane as
\begin{equation}
  \int_\Re d\phi_R \, \phi^n(\phi_R)\tilde{P}_\Gamma(\phi_R) = \frac{1}{Z_\Gamma(0)}\left.\frac{d^n Z_\Gamma(j)} {dj^n} \right|_{j=0},
\end{equation}
which can always be done by a real parametrization of the complex contour $\Gamma$, one sees that the equilibrium distribution can be written as a linear combination
\begin{equation}\label{Salcedo}
  \tilde{P}_{eq}(\phi_R) = \sum_{\Gamma_I} a_{\Gamma_I}\tilde{P}_{\Gamma_I}(\phi_R), 
\end{equation}
which is exactly the conjecture that was made by Salcedo \cite{Salcedo93}, where he derived the same result for a general complex distribution by considering the stationary solutions of the pseudo Fokker-Planck equation (\ref{Fokker_Planck_tilde}) to be realized as distributions rather than functions. Actually, we managed to refine his conjecture (which considers a sum over all $\Gamma$ instead of $\Gamma_I$ on the RHS of equation (\ref{Salcedo})) by showing that not all of $\tilde{P}_\Gamma(\phi_R)$ are independent through the use of Schwinger-Dyson equations. A final note is that the coefficients $a_\Gamma$ may depend on initial conditions. We will illustrate these points with numerical examples in the next section. \cite{Salcedo93} has other examples discussed along the lines mentioned here.

\subsection{Zero Dimensional Examples}

We consider two examples here, 
\begin{equation}
S_1(\phi)=i\frac{\phi^3}{3},\quad S_2(\phi)= -\frac{\phi^4}{4}.
\end{equation}
In both cases, we will derive an independent set of generating functionals and identify the particular solution of the Schwinger-Dyson equation to which the simulation converges by observing the sampling points in the complex plane. We will see that the change of initial conditions may change the resulting stationary distribution. Zero dimensional field theories have been heavily studied with complex Langevin equations before, e.g.\cite{Bernard01,Parisi83,Lin86,Okamoto89,Ito93,Hamber85,Flower86,Salcedo93}. What makes our presentation different from the previous studies is the relation to complex path integral solutions of Schwinger-Dyson equations.

In our simulations, we used the Euler method which is a first-order algorithm, see e.g. \cite{Gardiner04}. The complex Langevin equation (\ref{ComplexLangevin}) with this discretization is given by
\begin{align}
  \phi_R(\tau_{j+1}) &= \phi_R(\tau_j) -Re\left[ \frac{\partial S}{\partial \phi(\tau_j)}\right]\Delta\tau  + \sqrt{2\Delta\tau}\eta_j,\nonumber\\
\phi_I(\tau_{j+1}) &= \phi_I(\tau_j) -Im\left[ \frac{\partial S}{\partial \phi(\tau_j)}\right]\Delta\tau,\nonumber\\
& \tau_{j+1} = \tau_j + \Delta\tau, \qquad j \in \mathbb{Z},
\end{align}
where $\Delta\tau$ is the time step, and $\eta_j$ is a Gaussian random variable with zero mean and unit variance satisfying
\begin{equation}
  \left<\eta_j\right> = 0, \quad \left<\eta_j \eta_k \right> = \delta_{jk}.
\end{equation}
All our simulations run from $\tau_i=0$ to $\tau_f=1000$ with $\Delta\tau=0.001$. We start calculating expectation values after $\tau=5$. Error bars stand for the standard deviation of $50$ runs.

\begin{table}
\caption{\label{TableS1}Comparison of Langevin simulation results of action
  $S_1(\phi)$ with the correlators of the generating functions
  $Z_{1}^{(1)}(j)$ and $Z_{2}^{(1)}(j)$. The simulation ran from $\tau_i=0$ to
  $\tau_f=1000$ with $\Delta\tau=0.001$ with different initial conditions. We
  start calculating expectation values after $\tau=5$. Error bars stand for
  the standard deviation over $50$ runs. 5 diverging paths discarded for the
  second initial condition. Average is over 45 converging paths. See \cite{Ambjorn85} for a justification of this procedure. The last two rows show the corresponding exact values for the generating functions calculated by numerical integration.}
\begin{center}
\begin{tabular}{l|rrrr}
                 &$\left<\phi\right>$                      & $\left<\phi^2\right>$                   & $\left<\phi^3\right>$\\ \hline 
 \multirow{2}{*}{$\phi(0) = 0$}  & $-0.0034-i0.7289$& $i0.0053$ & $0.0025-i0.9998$\\ &$\pm0.0190\pm i0.0076$ & $\pm0.0003\pm i0.0364 $ & $\pm0.0292\pm i0.0347 $\\ \hline
 \multirow{2}{*}{$\phi(0) = 5i$} & $-0.0016-i0.7315$&$i0.0026$ & $0.0012-i1.0080$ \\  &$\pm0.0176\pm i0.0079$ & $\pm0.0003\pm i0.0325$ &  $\pm0.0260\pm i0.0330$\\ \hline 
  \multirow{2}{*}{$\phi(0) = 1-i$} & $-0.0049-i0.7289$ & $i0.0085$ & $0.0053-i0.9994$ \\ &$\pm0.0246\pm i0.0071$ & $\pm0.0003\pm i0.0457$ & $\pm0.0343\pm i0.0307$ \\ \hline
 
  $Z_{1}^{(1)}$  &  $-i0.7290$                            & $0$ & $-i$\\
  $ Z_{2}^{(1)}$ &  $-0.6313+i0.3645$                   & $0$ & $-i$\\  
\end{tabular}
\end{center}
\end{table}
\begin{figure}
\begin{center}
\includegraphics[width=140mm]{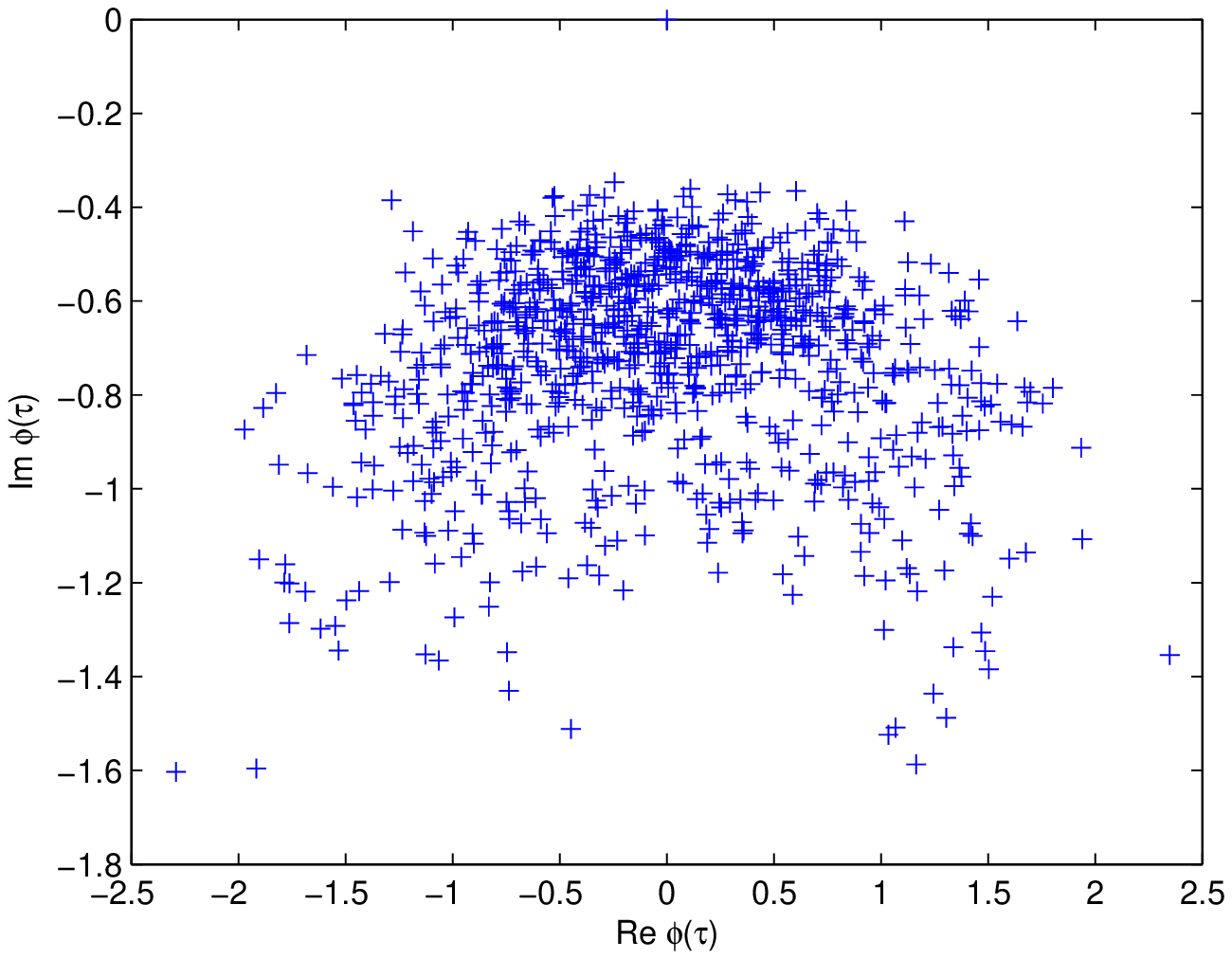}
\end{center}
\caption{\label{S1_sample} A sample path for the complex Langevin simulation of the action $S_1(\phi)$. Parameters are the same as in Table \ref{TableS1}. $1000$ sample points are shown with equal time intervals. Initial conditions is $\phi(0) = 0$.}
\end{figure}

For $S_1(\phi)$, an independent set of solutions to the Schwinger-Dyson equation (\ref{SD}) can be written by connecting the three zeros shown in Figure \ref{FigSectors}, i.e. $i\infty$, $e^{-i\frac{\pi}6}\infty$ and $e^{-i\frac{5\pi}{6}}\infty$. We choose the following generating functions $Z_{1}^{(1)}(j)$ and $Z_{2}^{(1)}(j)$: 
\begin{align}
  &Z^{(1)}_{1,2}(j) = \frac{\int_{\Gamma_{1,2}}d\phi \,\exp\left\lbrace- i\frac{\phi^3}{3}+j\phi\right\rbrace}{\int_{\Gamma_{1,2}}d\phi \,\exp\left\lbrace - i\frac{\phi^3}{3}\right\rbrace}, \nonumber \\
&\Gamma_1 = [e^{-i\frac{5\pi}{6}}\infty,0] + [0,e^{-i\frac{\pi}{6}}\infty], \nonumber \\ 
&\Gamma_2 = [e^{-i\frac{5\pi}{6}}\infty,0]+[0,i\infty].
\end{align}
We expect the result of the complex Langevin simulation to converge to a linear combination of the distributions defined by these generating functions. 

Table \ref{TableS1} shows the results of simulations for this theory. We compare with the exact results for the Green's functions of $Z_{1}^{(1)}(j)$ and $Z_{2}^{(1)}(j)$. Figure \ref{S1_sample} shows a sample path for the simulation of this action. We repeated the simulations with different initial conditions, some of which are given in Table \ref{TableS1}. For the converging simulations, the sample paths localized around the same region as of Figure \ref{S1_sample} and the obtained numerical values were similar to those of Table \ref{TableS1}. For some initial conditions with large $\left|\partial S/\partial\phi\right|$ values (e.g. $\phi(0)=5i$), we observed nonconverging paths, which could be made to converge (and localized in the same region of Figure \ref{S1_sample}) by decreasing the step size. This behavior can be understood by inspecting the deterministic part of the complex Langevin equation (i.e. without a noise term in equation (\ref{ComplexLangevin})). The solution to to the deterministic part will be:
\begin{equation}
  \phi_d(\tau)=\frac{\phi_d(0)}{1+i\phi_d(0)\tau},
\end{equation}
where $\phi_d(0)$ is the initial condition. We see that $\phi = 0$ is a global attractor for all points except the positive imaginary axis. Any path starting from the positive imaginary axis will go to infinity staying on the imaginary axis (i.e. $i\infty$) in finite time. When the noise term is included, which points along the real axis, these diverging paths will come out the positive imaginary axis and eventually approach the sampling region shown in Figure \ref{S1_sample}. However, numerically these paths may cause a problem. When the step size is not small enough, the simulation may go to infinity around these points in finite time. This is called the ``runaway solution'' problem, see \cite{Ambjorn85} for other examples and more details. Despite this numerical problem, which can be cured by smaller step sizes, the simulations suggest that for the action $S_1(\phi)$ the complex Langevin algorithm always converges. Inspecting Table \ref{TableS1} we see that the distribution defined by $Z_{1}^{(1)}(j)$ has correlators within the error range of numerical data. The sample path of Figure \ref{S1_sample} shows that the simulation does sample around the path of $Z_{1}^{(1)}(j)$. It definitely does not sample around the positive imaginary axis. Based on these observations, we conjecture that the complex Langevin simulation for $S_1(\phi)=i\frac{\phi^3}3$  theory always converge to the distribution defined by the generating function $Z_{1}^{(1)}(j)$.

\begin{table}
\caption{\label{TableS2}Comparison of Langevin simulation results of action $S_2(\phi)$ with correlators of the generating functions $Z_{a}^{(2)}(j)$, $Z_{b}^{(2)}(j)$ and $Z_{c}^{(2)}(j)$. The simulation ran form $\tau_i=0$ to $\tau_f=1000$ with $\Delta\tau=0.001$ with different initial conditions. We start calculating expectation values after $\tau=5$. Error bars stand for the standard deviation over $50$ runs. The last two rows show the corresponding exact values for the generating functions calculated by numerical integration.}
{\scriptsize
\begin{center}
\begin{tabular}{l|rrrr}
&$\left<\phi\right>$ & $\left<\phi^2\right>$ & $\left<\phi^3\right>$  & $\left<\phi^4\right>$ \\ \hline 
$\phi(0) = 0$  & no convergence & no convergence & no convergence & no convergence\\ \hline

 \multirow{2}{*}{$\phi(0) = i$} & $-0.0014+i0.9781$ & $-0.6765-i0.0029$ & $0.0038$ & $-1.0014-i0.0035$ \\  &$\pm0.0126\pm i0.0057$ & $\pm0.0078\pm i0.0295$ &  $\pm0.0452\pm i0.0003$& $\pm0.0462\pm i0.0242$\\ \hline 
 
\multirow{2}{*}{$\phi(0) = -i$}  & $-0.0008-i0.9794$ & $-0.6781+i0.0081$ & $0.0024$ & $-1.0054-i0.0041$ \\  &$\pm0.0138\pm i0.0071$ & $\pm0.0098\pm i0.0318$ &  $\pm0.0480\pm i0.0003$ & $\pm0.0312\pm i0.0491$\\ \hline

  $ Z_{a}^{(2)}$ &  $i0.9777$                            & $-0.6760$ & $0$  & $-1$\\
  $ Z_{b}^{(2)}$ &  $-i0.9777$                           & $-0.6760$ & $0$  & $-1$\\  
  $ Z_{c}^{(2)}$ &  $-0.9777$                            & $0.6760$  & $0$  & $-1$\\  
\end{tabular}
\end{center}
}
\end{table}
\begin{figure}
\begin{center}
\includegraphics[width=140mm]{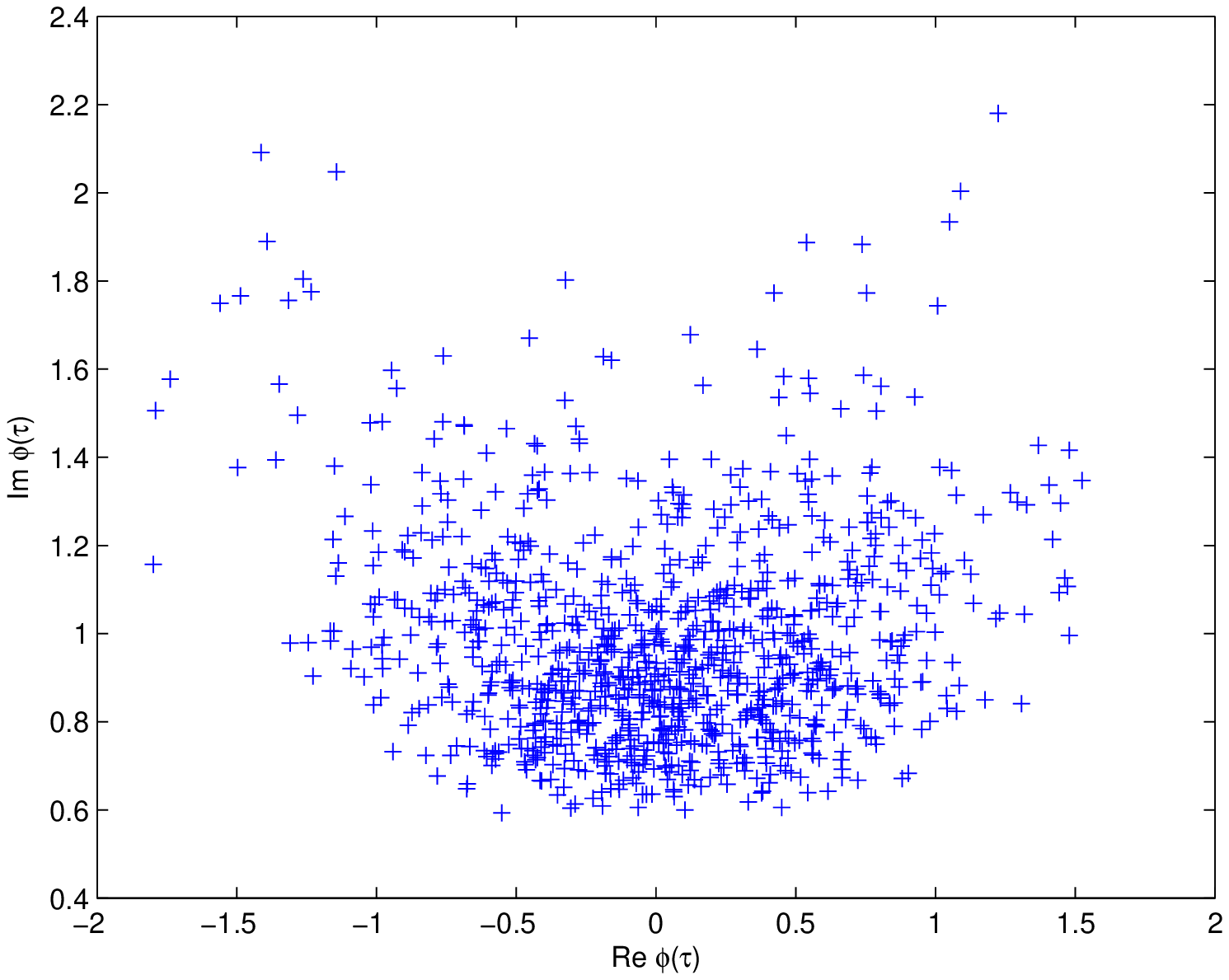}
\end{center}
\caption{\label{S2_sample_plus} A sample path for the complex Langevin simulation of the action $S_2(\phi)$. Parameters are the same as in Table \ref{TableS2}. $1000$ sample points are shown with equal time intervals. Initial conditions is $\phi(0) = i$.}
\end{figure}
\begin{figure}
\begin{center}
\includegraphics[width=140mm]{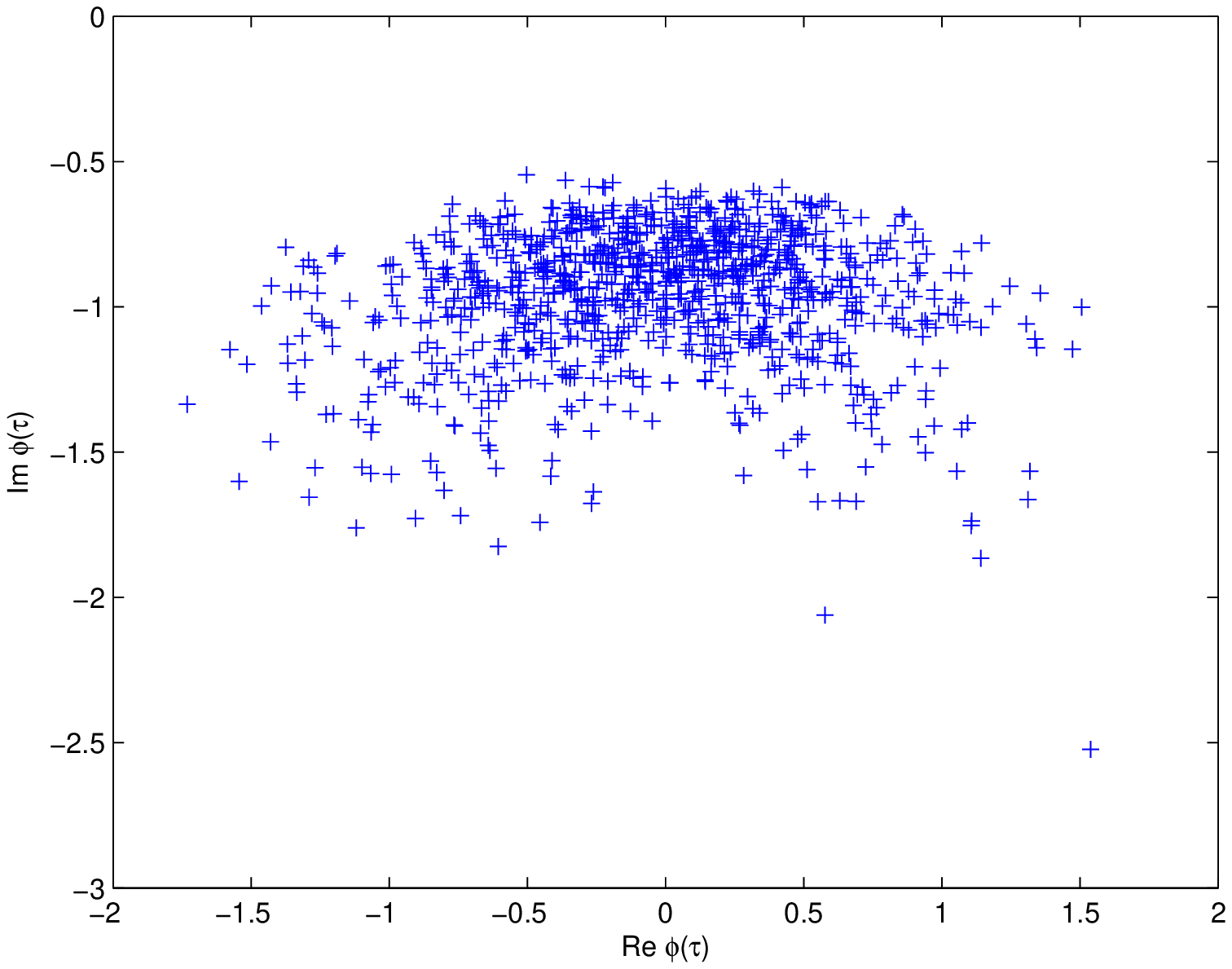}
\end{center}
\caption{\label{S2_sample_minus} A sample path for the complex Langevin simulation of the action $S_2(\phi)$. Parameters are the same as in Table \ref{TableS2}. $1000$ sample points are shown with equal time intervals. Initial conditions is $\phi(0) = -i$.}
\end{figure}

Now we turn to $S_2(\phi) = -\phi^4/4$. Note that this action is real, however $e^{-S(\phi)}$ does not define a probability distribution on the real line, as the integral is divergent. The complex evolution of the associated Langevin equation is introduced by a choice of complex initial conditions. This procedure, choosing complex initial conditions, will turn every real Langevin process to a complex Langevin process. The statements we made about complex Langevin equations are valid for these cases also.

An independent set of generating functions for this theory can be written by connecting the four zeros of $S_2(\phi)$ on the complex plane, i.e. $e^{i\frac{\pi}4}\infty$,  $e^{-i\frac{\pi}4}\infty$,  $e^{i\frac{3\pi}4}\infty$ and  $e^{-i\frac{3\pi}4}\infty$. We choose the following generating functions $Z_{1}^{(2)}(j)$, $Z_{2}^{(2)}(j)$ and $Z_{3}^{(2)}(j)$: 
\begin{align}
  &Z^{(2)}_{a,b,c}(j) = \frac{\int_{\Gamma_{a,b,c}}d\phi \,\exp\left\lbrace\frac{\phi^4}{4}+j\phi\right\rbrace}{\int_{\Gamma_{a,b,c}}d\phi \,\exp\left\lbrace \frac{\phi^4}{4}\right\rbrace}, \nonumber \\
&\Gamma_a = [e^{i\frac{3\pi}{4}}\infty,0] + [0,e^{i\frac{\pi}{4}}\infty], \nonumber \\ 
&\Gamma_b = [e^{-i\frac{3\pi}{4}}\infty,0]+[0,e^{-i\frac{\pi}{4}}\infty], \nonumber \\ 
&\Gamma_c = [e^{-i\frac{3\pi}{4}}\infty,0]+[0,e^{i\frac{3\pi}{4}}\infty].
\end{align}
We expect to see the complex Langevin simulation converge to a linear combination of the distributions defined by these generating functions.

Table \ref{TableS2} shows the simulation results with different initial conditions compared with the correlators of the generating functions. In contrast to the previous case, we see that the results of the simulation is initial value dependent. Initial points on the real axis do not converge at all. Initial values above and below the real axis converge, but the sample points localize in different regions of the complex $\phi$ plane giving different results, see Figures \ref{S2_sample_plus} and \ref{S2_sample_minus}. We can understand this behavior again by inspecting the deterministic part of the complex Langevin equation. This time the solution will be: 
\begin{equation}
  \phi_d(\tau)= \frac{\phi_d(0)}{\sqrt{1-2[\phi_d(0)]^2\tau}}         
\end{equation}
where $\phi_d(0)$ is the initial condition and the square root function gives the principal root. We see that $\phi=0$ is a global attractor for every point on the complex plane except the real line. The origin will repell any path starting on the real line; these paths will diverge in finite time. When the real noise term is added, simulations with real initial points will stay on the real line and due to the repulsion they will diverge. Initial points on the upper/lower half of the complex plane will be attracted by the origin and the simulations will sample in the upper/lower half plane as seen in  Figures \ref{S2_sample_plus} and \ref{S2_sample_minus}. As a result, upper half plane initial points will converge to a different solution of the Schwinger-Dyson equation than those of the lower half plane initial points. Table \ref{TableS2} shows two simulations starting with $\phi(0)=i$ and $\phi(0)=-i$. We see that the correlators of $Z_a^{(2)}$ are in the error range of the former initial condition, while the correlators of $Z_b^{(2)}$ are in the range of the latter simulation. Based on these observations, we conjecture that for $S_2(\phi)$, the complex Langevin equation will diverge if the initial condition is real, converge to the distribution defined by $Z_a^{(2)}(j)$ if the initial condition is on the upper half of the complex plane or converge to the distribution defined by $Z_b^{(2)}(j)$ otherwise. 

One might suspect that either or both of the two stationary states we discussed are quasi-stationary states and consequently expect to see the system converge to a unique stable stationary state after a long enough simulation time. Assuming convergence to a stationary state, we will argue that this is not the case, initial points above and below the real line will behave differently in the whole range of simulation. Since the path of the complex field $\phi$ is continuous (but not differentiable), a complex Langevin process starting from above the real line will never end up below the real line. The reverse statement holds also. The reason is that any path going from one half plane to the other must pass through the real line, and once the path is on the real line, it stays on the real line. Both the noise term and the drift term points along the real line. Furthermore, the path on the real line will show nonconvergent behavior as discussed above. So either, all paths diverge at the end, or, assuming convergence, initial point above and below the real line end up sampling in different regions of the complex plane, converging to different stationary distributions. This initial value dependence is crucial in complex Langevin equations.

Actions $S_1(\phi)$ and $S_2(\phi)$ were studied in the context of
$\cal{PT}$-symmetric quantum field theories in \cite{Bernard01}, where only
one of the path integral solutions to the Schwinger-Dyson equation is
considered, e.g. \cite{Bender06}. In fact, the authors of \cite{Bernard01},
Bernard and Savage, provided a formal proof that the complex Langevin equation
for $S_1(\phi)$ should always converge to the distribution defined by
$Z_1^{(1)}(j)$, regardless of the initial condition, which we have also
observed in our simulations. The formal proof is based on the methods of
\cite{Nakazato87} and involves defining $\tilde{P}(\phi_c,\tau)$ of equation
(\ref{tilde}) as a projection of $P(\phi_R,\phi_I,\tau)$ to the complex
integration contour $\Gamma_1$ (after necessary partial integrations), as
opposed to the real line as of equation (\ref{P_tilde}). Then the associated
Fokker-Planck Hamiltonian will have a real and positive spectrum with one zero
eigenvalue. Bernard and Savage argued that with slight modifications the same
reasoning applies for $S_2(\phi)$ with the integration contour of
$\Gamma_b$. They also noted that in simulations of $S_2(\phi)$ one has to
choose the initial point $\phi(0)$ to be in the lower half of the complex
$\phi$ plane or else the numerical simulations will be unstable. In our
studies, we observed instabilities for initial points on the real line, which
we interpreted to be nonconvergent behavior. Initial points in the lower and
upper halves of the complex $\phi$ plane were observed to converge to
different probability distributions (solutions of the Schwinger-Dyson
equation). Our results suggest that the proof given by Bernard and Savage
should be modified to include the effects of different initial conditions and
all possible solutions of the Schwinger-Dyson equation. We believe that more
 attention must be paid to the boundary conditions of $P(\phi_R,\phi_I,\tau)$
when doing the projection.

Another point to note is that in both cases, we could not recover the whole solution set of the Schwinger-Dyson equations from the complex Langevin equation. In general, $\phi^N$ theories will have $N-1$ independent solutions to the Schwinger-Dyson equations. In the cases that we studied, we could recover only $N-2$ of them. We may need to consider other stochastic processes to recover the whole solution set.

\section{Lattice Study}\label{lattice}

It is trivial to generalize this discussion to a lattice. In particular, consider a general Euclidean scalar field theory with a polynomial potential term,
\begin{align}\label{latact}
\mathcal{L} = \partial^2\phi(x) + \sum_{l=1}^{n}\frac{g_l}{l}\phi^l(x). 
\end{align}
For simplicity we consider a lattice in one dimension, which we call time and
denote by $t$. The generating functional of the theory is now a function of
$m$ variables, where $m$ is the number of lattice points. We denote the
generating functional by $Z(j)=Z(j_1,\ldots,j_m)$. $j_i$ is the source at
$i^{\text{th}}$ lattice point. We will use $j=0$ to mean $j_1=\ldots=j_{m}=0$. By definition, correlation functions are given
by
\begin{align}
\left<\phi_1^{k_1}\ldots\phi_{m}^{k_{m}}\right> =
\left.\frac{\partial^{k_1}}{\partial j_1^{k_1}}\ldots\frac{\partial^{k_{m}}}{\partial j_{m}^{k_{m}}}Z(j)\right|_{j=0}.
\end{align}

Next, we introduce the difference operators on the lattice,
\begin{align}
\Delta_+ = \phi_{i+1}-\phi_{i-1}, \qquad \Delta_- = \phi_{i} - \phi_{i-1},
\end{align}
where $\phi_i$ stands for the field value at $i^{\text{th}}$ lattice
point. The Schwinger-Dyson equation for the scalar field theory \eqref{latact}
can be written by a coupled set of partial differential equations. Using a
centered discretization for time derivatives, the system is composed of a partial differential equation for every lattice point, given by
\begin{align}\label{latSD}
\left(\Delta_+\Delta_-\frac{\partial}{\partial j_i}+\sum_{l=1}^ng_l\frac{\partial^{l-1}}{\partial j^{l-1}_i}\right)Z(j) = j_iZ(j),
\end{align}
where the lattice spacing is set to one. 

This discretization requires one to set boundary conditions on $Z(j)$ at
initial and final lattice points. There are many possible choices,
we choose periodic boundary conditions. We assume that the whole space is filled
with a lattice of period $m$.

In the absence of the kinetic term, each lattice point acts
independently. Therefore, study of the complex Langevin equation for zero
dimensional case immediately tells us what will happen on the lattice. When
the kinetic term is included, couplings between lattice points take action and
problem is more complicated.

A general solution to the lattice Schwinger-Dyson equation can be written as
\begin{align}\label{LatGen}
Z(j) = \frac 1{\mathcal{N}} \int_{\Gamma_1} d\phi_1\ldots &\int_{\Gamma_m}
d\phi_m \nonumber \\
&\exp\left\lbrace-\sum_{i=1}^m\left(\frac 12 \phi_i\Delta_+\Delta_-\phi_i +
\sum_{l=1}^n\frac{g_l}{l}\phi_i^l\right)+\sum_{i=0}^{m+1}j_i\phi_i\right\rbrace,
\end{align}
where $\Gamma_i$ denote contours that connect the zeros of the integrand on
complex $\phi_i$ plane. We again normalize so that $Z(0)=1$. For
each lattice point there are $n$ independent solutions, which leads to
$m\times n$ independent solutions to the whole lattice Schwinger-Dyson
equation. Because of linearity any combination of these solutions is also a
solution. We note that there is no ambiguity in the solution written in this form, one can do the integrations in any order. 

We turn to the complex Langevin equation for this problem. We again introduce
a fictitious time coordinate $\tau$. The complex Langevin system is now
written in terms of stochastic variables $\phi_j(\tau)=\phi_{Rj}(\tau) + i
\phi_{Ij}(\tau)$. For each lattice point $j = 1,\ldots,m$, there is a
stochastic equation:
\begin{align}
d\phi_j(\tau) = -\Delta_+\Delta_-\phi_j(\tau) - \sum_{l=1}^n g_l\phi_j(\tau)^l +dw_j(\tau),
\end{align}
where $dw_j(\tau)$ are $m$ independent Wiener processes normalized as
before. A repetition of the analysis of the previous section is sufficient to conclude that
the stationary distributions of this set of stochastic equations will satisfy
lattice Schwinger-Dyson equations \eqref{latSD}. 

Let's consider again the $-\phi^4/4$ theory, this time in one dimension. The
Lagrangian is given by
\begin{equation}
\mathcal{L} = \frac 12 \left(\frac{\d\phi}{dt}\right)^2 - \frac g4 \phi^4.
\end{equation}
We again note that this theory is bottomless, a normal (real line contour)
path integral solution to the Schwinger-Dyson equations does not
exist. However, complex contour contour solutions do exist. Table \ref{TableLat}
shows the results of numerical simulations for this theory on a one
dimensional lattice. As in zero dimensional case, we see that the complex Langevin
equation has at least two different stationary distributions. Choice of
initial conditions can alter the stationary distributions. We note that this
theory was also studied in \cite{Bernard01} and there initial conditions were
restricted to lower half of the complex plane. This led to the observation of only one of the stationary distributions, which was concluded to be the stationary distribution that led to a $\mathcal{PT}$-symmetric theory. This could be described by choosing $\Gamma_b$ of the previous section at each lattice point as the integration contour for the generating functional \eqref{LatGen}. Table \ref{TableLat} suggests that the other solution is given by choosing $\Gamma_a$ at every point.

\begin{table}
\caption{\label{TableLat} A complex Langevin simulation is done for the one
  dimensional $-\phi^4/4$ theory. 128 lattice points are taken with spacing
  set to 1. An Euler method is used for fictitous time evolution. Simulations
  ran from $\tau_i=0$ to $\tau_f=10000$, with $\Delta\tau = 0.001$. The table
  below lists the first four equal time correlation functions obtained for two
different initial conditions. We observed translation symmetry on the equal
time correlation functions as expected. We list average values over lattice
points, e.g. $\left<\phi\right> = \frac
1{128}\sum_{i=1}^{128}\left<\phi_i\right>$. Standard deviations are calculated
over lattice points. We also considered other initial conditions, which
resulted in either nonconvergence or convergence to one of the two cases shown
below. A detailed study of initial condition dependence is beyond the scope of this paper.}
{\scriptsize
\begin{center}
\begin{tabular}{l|rrrr}
&$\left<\phi\right>$ & $\left<\phi^2\right>$ & $\left<\phi^3\right>$  & $\left<\phi^4\right>$ \\ \hline 
 \multirow{1}{*}{$\phi(0,t) = -i$} & $-i0.8891$ & $-0.5537+i0.0001$ &
 $0.0001$ & $-0.6307-i0.0001$ \\  &$\pm0.0032\pm i0.0012$ & $\pm0.0013\pm
 i0.0065$ &  $\pm0.0083\pm i0.0018$& $\pm0.0069\pm i0.0063$\\ \hline 

 \multirow{2}{*}{$\phi(0,t) = i$} & $i0.8891$ & $-0.5537-i0.0001$ & $0.0001$ & $-0.6306+i0.0001$ \\  &$\pm0.0037\pm i0.0013$ & $\pm0.0013\pm i0.0075$ &  $\pm0.0095\pm i0.0021$& $\pm0.0071\pm i0.0071$\\ \hline 
\end{tabular}
\end{center}
}
\end{table}

More studies on the lattice must be done to understand the convergence
behavior of complex Langevin equations. Here, as well as initial conditions, boundary conditions on the lattice may also take effect. Some other specific questions are listed in
the next section. We leave detailed lattice studies to a coming paper. The point of this section is to demonstrate that the results of the previous section applies to the lattice as well.

\section{Discussion and Conclusion}

Salcedo \cite{Salcedo93} suggested that stationary distributions of the complex Langevin equation may be interpreted as different phases of the associated quantum field theory. Some authors used Langevin and complex Langevin equation to give physical meaning to bottomless actions, e.g. \cite{Greensite84,Okamoto89,Tanaka92,Tanaka93,Ito93}\footnote[1]{Some of these references use nonconstant kernels in the complex Langevin equation which enlarges the set of stationary distributions \cite{Salcedo93} and changes the Schwinger-Dyson equation, e.g. \cite{Ito93}}. Our discussion shows that a rephrasal of these questions is to ask which solutions of the Schwinger-Dyson equations define a phase of quantum field theory. Here we discuss this point, and the relation of different phases to the boundary conditions of Schwinger-Dyson equations \cite{Garcia96,Guralnik07}. 

Schwinger-Dyson equations are differential equations and admit more than one
 solution. Therefore it is necessary to set boundary conditions to specify the
particular solution one is looking for. If one is solving for a quantum field
theory using Schwinger-Dyson equations, it seems reasonable to choose the
boundary condition so that the solution is the standard path integral over
real fields. However, in many cases this solution actually will not be the
physical one, e.g. symmetry breaking phases. Also, in theories with actions
unbounded below, the integrals over real fields are not even convergent. In
these cases, it is reasonable to look at other solutions of the
Schwinger-Dyson equations and study them as possible generating functionals of
the associated quantum field theory. Of course, different solutions require
specification of different boundary conditions. One way of specifying
different solutions is to consider path integrals over complex paths (as
opposed to real paths) that connect zeros of the partition function on the
complex plane, as was demonstrated in equation (\ref{contour}). Some of these
also happen to be the stationary distributions of the associated complex
Langevin equation constructed by Salcedo \cite{Salcedo93} as shown in this
paper. The problem with this approach is the large number of different
boundary conditions/solutions and if all these different solutions define a
phase or vacuum of the associated quantum field theory. A possible reduction
of the solution set comes from taking the thermodynamic limit of the
lattice. These issues are discussed in detail in
\cite{Garcia96,Guralnik07}.

In the light of the discussion above, one concludes that to study the phase structure of quantum field theories, one needs to study the different solutions to Schwinger-Dyson equations. This task requires new numerical methods in the study of quantum field theory. One suggestion is the Source-Galerkin method, see e.g. \cite{Dmitri05} and references therein. This method proposes an expansion of the generating functional in polynomials of the source term and optimizes this expansion by a Galerkin procedure using the Schwinger-Dyson equation. It is successful in many problems, but also proved to be very difficult in many other cases. Another approach is given by mollification of the path integral weight \cite{Ferrante06}. The connection between complex Langevin and Schwinger-Dyson equations suggests the use of complex Langevin simulations as a numerical method to study the phase structure of quantum field theories.

Some questions and speculations in this quest are:
\begin{itemize}
\item Is it possible to know a priori if the complex Langevin equation will converge?

\item What is the exact relation between the initial condition and the stationary distributions of the complex Langevin equation?

\item Is it possible to recover the whole solution set of the Schwinger-Dyson equation using the complex Langevin equation? If not, can we use other stochastic systems to recover the whole set? Also, what is special about the recovered solutions?

\item How are these results modified in the continuum? There are an infinite number of solutions to Schwinger-Dyson equations on the lattice. The continuum limit may cause a collapse in the solution set. This is definitely the case if space time translational invariance is required. \cite{Garcia96,Guralnik07}. Can one see this collapse using complex Langevin equations? This will be very important in understanding the phase structure of quantum field theories.
\end{itemize}
We address some of these problems in future work.

\section*{Acknowledgments}
The authors would like to thank D. D. Ferrante and D. Obeid for useful discussions and conversations. G. Guralnik would like to thank Z. Guralnik for many conversations related to this work. This work is supported in part by funds provided by the US Department of Energy (DoE) under DE-FG02-91ER40688-TaskD.

\bibliographystyle{elsart-num}
\bibliography{CL}

\end{document}